%% file: paper.tex
\documentclass[11pt,a4paper]{article}
\usepackage[hyperref]{acl2020}
\usepackage{CJKutf8}
\usepackage[utf8]{inputenc}

\usepackage{times}
\usepackage{latexsym}

\usepackage{graphicx}

\usepackage{subfigure}
\usepackage{mathrsfs}
\usepackage{paralist}
\usepackage{amsthm}
\usepackage{enumitem}
\usepackage{enumerate}
\usepackage{marvosym}

\usepackage{xspace}
\usepackage{paralist}

\usepackage{microtype}

\aclfinalcopy 
\def\aclpaperid{35} 

\newcommand{\method}{Xiaomingbot\xspace}

\begin{document}

\title{Xiaomingbot: A Multilingual Robot News Reporter}

\author{Runxin Xu$^{1}$\Thanks{ The work was done while the author was an intern at ByteDance AI Lab.}, Jun Cao$^2$, Mingxuan Wang$^2$, Jiaze Chen$^2$, Hao Zhou$^2$, Ying Zeng$^2$, Yuping Wang$^2$ \\
\textbf{Li Chen}$^2$, \textbf{Xiang Yin}$^2$, \textbf{Xijin Zhang}$^2$, \textbf{Songcheng Jiang}$^2$, \textbf{Yuxuan Wang}$^2$,  \and \textbf{Lei Li}$^2$\Thanks{ Corresponding author.} \\
$^1$ School of Cyber Science and Engineering, Shanghai Jiao Tong University, Shanghai, China \\
$^2$ ByteDance AI Lab, Shanghai, China \\
\texttt{runxinxu@gmail.com} \\
\texttt{ \{caojun.sh, wangmingxuan.89, chenjiaze, zhouhao.nlp, zengying.ss, } \\
\texttt{ wangyuping, chenli.cloud, yinxiang.stephen, zhangxijin,} \\
\texttt{ jiangsongcheng, wangyuxuan.11, lileilab\}@bytedance.com} 
}

\date{}

\maketitle

\begin{abstract}
    \input{00abs.tex}

\end{abstract}

\section{Introduction}
\label{sec:intro}
\input{01intro.tex}

\section{System Architecture}
\label{sec:system}

\input{02frame.tex}

\section{News Generation}
\label{sec:news-gen}
\input{03xiaoming.tex}

\section{News Translation}
\label{sec:news-trans}
\input{035news-trans.tex}

\section{Multilingual News Reading}
\label{sec:tts}
\input{04tts.tex}

\section{Synchronized Avatar Animation Synthesis}
\label{sec:avatar}

\input{05avatar.tex}

\section{Conclusion}
\label{sec:conclusion}
\input{06conclusion.tex}

\subsubsection*{Acknowledgments}
We would like to thank Yuzhang Du, Lifeng Hua, Yujie Li, Xiaojun Wan, Yue Wu,  Mengshu Yang, Xiyue Yang, Jibin Yang, and Tingting Zhu for helpful discussion and design of the system. The name Xiaomingbot was suggested by Tingting Zhu in 2016. 
We also wish to thank the reviewers for their insightful comments.

\bibliography{acl2020}
\bibliographystyle{acl_natbib}

\end{document}

%% file: 00abs.tex
    This paper proposes the building of \method, an intelligent, multilingual and multimodal software robot equipped with four integral capabilities: news generation, news translation, news reading and avatar animation. Its system summarizes Chinese news that it automatically generates from data tables. Next, it translates the summary or the full article into multiple languages, and reads the multilingual rendition through synthesized speech.  Notably, \method utilizes a voice cloning technology to synthesize the speech trained from a real person's voice data in one input language.
    The proposed system enjoys several merits: it has an animated avatar, and is able to generate and read multilingual news. Since it was put into practice, \method has written over 600,000 articles, and gained over 150,000 followers on social media platforms.

%% file: 01intro.tex

The wake of automated news reporting as an emerging research topic has witnessed the development and deployment of several robot news reporters with various capabilities. Technological improvements in modern natural language generation have further enabled automatic news writing in certain areas.
For example, GPT-2 is able to create fairly plausible stories~\cite{radford2019language}.
Bayesian generative methods have been able to create descriptions or advertisement slogans from structured data~\cite{miao2019cgmh,ye2020variational}.
Summarization technology has been exploited to produce reports on sports news from human commentary text~\cite{zhang-etal-2016-towards}.  

While very promising, most previous robot reporters and machine writing systems have 
limited capabilities reports on sports news that only focus on text generation. 
We argue in this paper that an intelligent robot reporter should acquire the following capabilities to be truly user friendly:
\begin{inparaenum}[\it a)]
    \item it should be able to create news articles from input data;
    \item it should be able to read the articles with lifelike character animation like in TV broadcasting; and
    \item it should be multi-lingual to serve global users. 
\end{inparaenum}
None of the existing robot reporters are able display performance on these tasks that matches that of a human reporter. 
In this paper, we present \method, a robot news reporter  capable of news writing, summarization, translation, reading, and visual character animation.
In our knowledge, it is the first multilingual and multimodal AI news agent. Hence, the system shows great potential for large scale industrial applications.

\begin{figure}[!t]
    \centering
    \includegraphics[width=0.5\textwidth,keepaspectratio,trim={5.6cm 7.5cm 14cm 10.2cm}, clip]{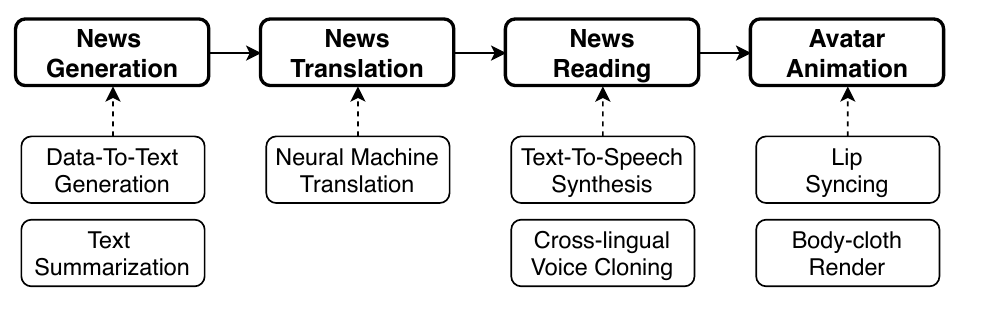}
    \caption{\method System Architecture}
    \label{fig:system-diagram}
\end{figure}

\begin{figure*}[ht]
    \centering
    \includegraphics[width=1.0\textwidth,keepaspectratio]
    {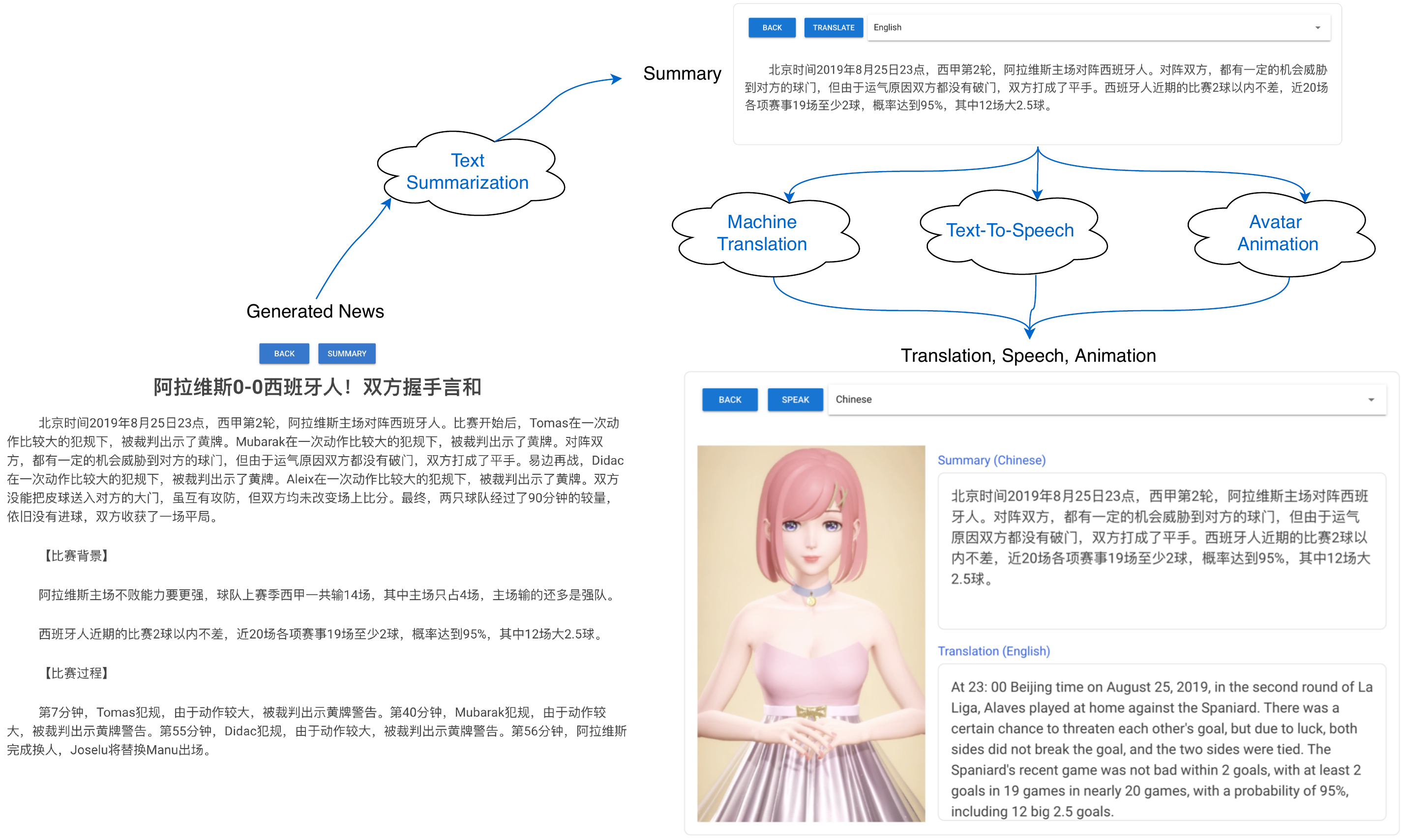}
    \caption{User Interface of \method. On the left is a piece of sports news, which is generated from a Table2Text model. On the top is the text summarization result. On the bottom right corner, \method produces the corresponding speech and visual effects.}
    \label{fig:ui-overview}
\end{figure*}

Figure~\ref{fig:system-diagram} shows the capabilities and components of the proposed \method system. It includes four components: 
\begin{inparaenum}[\it a)]
    \item a news generator, 
    \item a news translator,
    \item a cross-lingual news reader, and 
    \item an animated avatar.
\end{inparaenum}
The text generator takes input information from data tables and produces articles in natural languages. 
Our system is targeted for news area with available structure data, such as sports games and financial events. 
The fully automated news generation function is able to write and publish a story within mere seconds after the event took place, and is therefore much faster compared with manual writing. 
Within a few seconds after the events, it can accomplish the writing and publishing of a story.
The system also uses a pretrained text summarization technique to create summaries for users to skim through. 
\method can also translate news so that people from different countries can promptly understand the general meaning of an article. 
\method is equipped with a cross lingual voice reader that can read the report in different languages in the same voice.
It is worth mentioning that \method excels at voice cloning. It is able to learn a person's voice from audio samples that are as short as only two hours, and maintain precise consistency in using that voice even when reading in different languages. 
In this work, we recorded 2 hours of Chinese voice data from a female speaker, and \method learnt to speak in English and Japanese with the same voice.
Finally, the animation module produces an animated cartoon avatar with lip and facial expression synchronized to the text and voice.
It also generates the full body with animated cloth texture. 
The demo video is available at \url{https://www.youtube.com/watch?v=zNfaj_DV6-E}. The home page is available at \url{https://xiaomingbot.github.io}.

The system has the following advantages:
\begin{inparaenum}[\it a)]
    \item It produces timely news reports for certain areas and  is multilingual.
    \item By employing a voice cloning model to Xiaomingbot's neural cross lingual voice reader, we've allowed it to learn a voice in different languages with only a few examples 
    \item For better user experience, we also applied cross lingual visual rendering model, which generates synthesis lip syncing in consistent  with the generated voice.
    \item \method has been put into practice and produced over $600,000$ articles, and gained over 150k followers in social media platforms.
\end{inparaenum}

%% file: 02frame.tex
The \method system includes four components working together in an pipeline, as shown in Figure~\ref{fig:system-diagram}. 
The system receives input from data table containing event records, which, depending on the domain, can be either a sports game with time-line information, or a financial piece such as tracking stock market. The final output is an animated avatar reading the news article with a synthesized voice.
Figure \ref{fig:ui-overview} illustrates an example of our \method system.
First, the text generation model generates a piece of sports news. Then, as is shown on the top of the figure, the text summarization module trims the produced news into a summary, which can be read by users who prefer a condensed abstract instead of the whole news. Next, the machine translation module will translate the summary into the language that the user specifies, as illustrated on the bottom right of the figure. Relying on the text to speech (TTS) module, \method can read both the summary and its translation in different languages using the same voice. Finally, the system can visualize an animated character with synchronized lip motion and facial expression, as well as lifelike body and clothing. 

%% file: 03xiaoming.tex
In this section, we will first describe the automated news generation module, followed by the news summarization component.

\subsection{Data-To-Text Generation}
Our proposed \method is targeted for writing news for domains with structured input data, such as sports and finance. 
To generate reasonable text, several methods have been proposed\cite{miao2019cgmh,sun2019graspsnooker,ye2020variational}. 
However, since it is difficult to generate correct and reliable content through most of these methods, we employ a  template based on table2text technology to write the articles.

\begin{CJK}{UTF8}{gbsn}
\begin{table*}[htb]
    \centering
    \caption{Examples of Sports News Generation}
    \begin{tabular}{llll|p{1.6in}|p{1.6in}}
         \hline
         Time& Category & Player & Team & Generated Text & Translated Text \\
         \hline
         23’& Score & Didac & Espanyol & 第23分钟，西班牙人迪达克打入一球。 & In the 23rd minute, Espanyol Didac scored a goal. \\
         \hline
         35'&Yellow Card & Mubarak & Alav\'{e}s& 第35分钟，阿拉维斯穆巴拉克吃到一张黄牌。& In the 35th minute, Alav\'{e}s Mubarak received a yellow card.\\
         \hline
    \hline
    \end{tabular}
    \label{tab:data2text-example}
\end{table*}
\end{CJK}

Table~\ref{tab:data2text-example} illustrates one example of soccer game data and its generated sentences. 
In the example, Xiaomingbot retrieved the tabled data of a single sports game with time-lines and events, as well as statistics for each player's performance. 
The data table contains time, event type (scoring, foul, etc.), player, team name, and possible additional attributes.
Using these tabulated data, we integrated and normalized the key-value pair from the table. 
We can also obtain processed key-value pairs such as ``Winning team'', ``Lost team'', ``Winning Score''
, and use  template-based method to generate news from the tabulated result.
Those templates are written in a custom-designed java-script dialect. 
For each type of the event, we manually constructed multiple templates and the system will randomly pick one during generation. 
We also created complex templates with conditional clauses to generate certain sentences based on the game conditions. 
For example, if the scores of the two teams differ too much, it may generate ``Team A overwhelms Team B.''
Sentence generation strategy are classified into the following categories:
\begin{itemize}
\item \textbf{Pre-match Analysis.} It mainly includes the historical records of each team.
\item \textbf{In-match Description.} It describes most important events in the game such as ``someone score a goal'', ``someone received yellow card''.
\item \textbf{Post-match Summary.} It's a brief summary of this game , while also including predictions of the progress of the subsequent matches.
\end{itemize}

\subsection{Text Summarization}
For users who prefer a condensed summary of the report, \method can provide a short gist version using a pre-trained text summarization model. We choose to use the said model instead of generating the summary directly from the table data because the former can create more general content, and can be employed to process manually written reports as well. There are two approaches to summarize a text: extractive and abstractive summarization.
Extractive summarization trains a sentence selection model to pick the important sentences from an input article, while an
abstractive summarization will further rephrase the sentences and explore the potential for combining multiple sentences into a simplified one. 

We trained two summarization models. 
One is a general text summarization using a BERT-based sequence labelling network. 
We use the TTNews dataset, a Chinese single document summarization dataset for training from NLPCC 2017 and 2018 shared tasks~\cite{hua2017overview,li2018overview}.
It includes 50,000 Chinese documents with human written summaries.
The article is separated into a sequence of sentences. 
The BERT-based summarization model output 0-1 labels for all sentences. 

In addition, for soccer news, we trained a special summarization model based on the commentary-to-summary technique~\cite{zhang-etal-2016-towards}.
It considers the game structure of soccer and handles important events such as goal kicking and fouls differently. 
Therefore it is able to better summarize the soccer game reports.

%% file: 035news-trans.tex
In order to provide multilingual news to users, we propose using a machine translation system to translate news articles. In our system, we pre-trained several neural machine translation models, and employ state of the art Transformer Big Model as our NMT component. The parameters are exactly the same with~\cite{vaswani2017attention}. 
In order to further improve the system and speed up the inference, we implemented a CUDA based NMT system, which is 10x faster than the Tensorflow approach~\footnote{https://github.com/bytedance/byseqlib}.
Furthermore, our machine translation system leverages named-entity (NE) replacement for glossaries including team name, player name and so on to improve the translation accuracy.
It can be further improved by recent machine translation techniques   \cite{yang2020towards,Zheng2020Mirror-Generative}.

\begin{figure}[htb]
    \centering
    \includegraphics[width=0.5\textwidth]{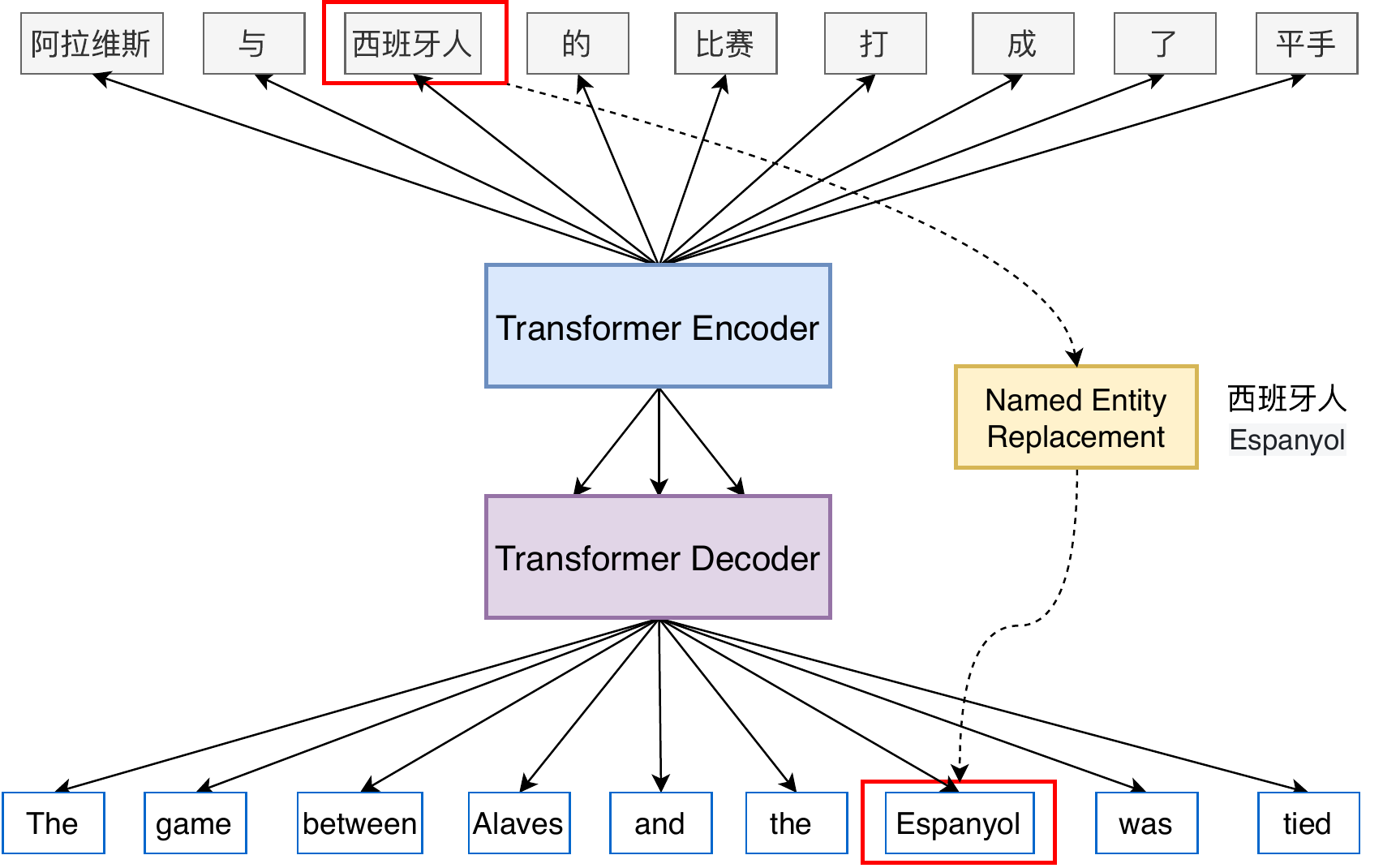}
    \caption{Neural Machine Translation Model.}
    \label{fig:NMT}
\end{figure}

We use the in-house data to train our machine translation system.
For Chinese-to-English,  the dataset contains more than 100 million parallel sentence pairs.   
For Chinese-to-Japanese, the dataset contains more than 60 million parallel sentence pairs.

%% file: 04tts.tex
\begin{figure}[htb]
    \centering
    \includegraphics[width=10cm,keepaspectratio,trim={9.5cm 1cm 10cm 8.5cm}, clip]{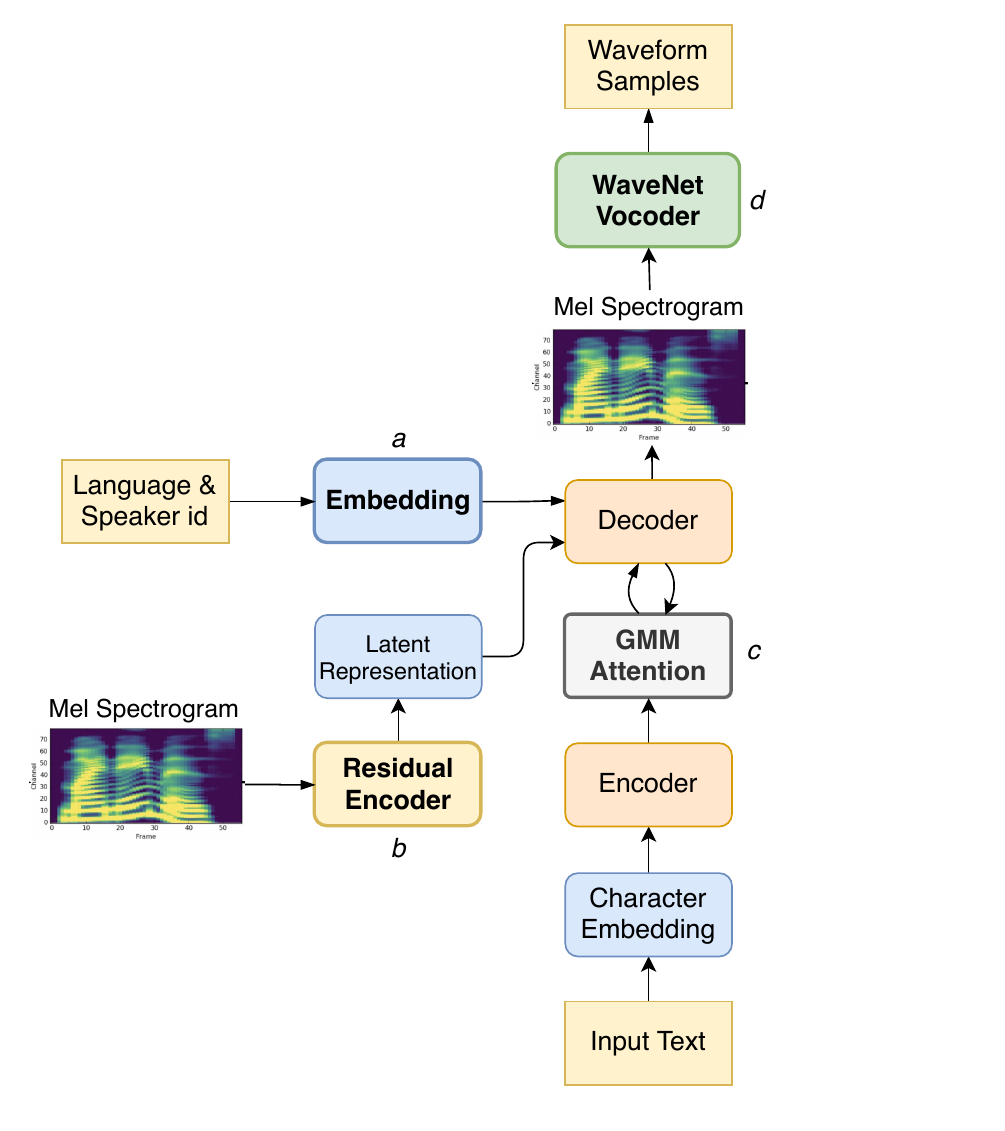}
    \caption{Voice Cloning for Cross-lingual Text-to-Speech Synthesis.}
    \label{fig:crosslingual-tts}
\end{figure}

In order to read the text of the generated and/or translated news article, we developed a text to speech synthesis model with multilingual capability, which only requires a small amount of recorded voice of a speaker in one language.
We developed an additional cross-lingual voice cloning technique to clone the pronunciation and intonation. 
Our cross-lingual voice cloning model is based on Tacotron 2~\cite{J.Shen2018:icassp}, which uses an attention-based sequence-to-sequence model to generate a sequence of log-mel spectrogram frames from an input text sequence~\cite{wang2017tacotron}. The architecture is illustrated in Figure~\ref{fig:crosslingual-tts}, we made the following augmentations on the base Tacotron 2 model:
\begin{itemize}
\item We applied an additional speaker as well as language embedding to support multi-speaker and multilingual input.
\item We introduced a variational autoencoder-style residual encoder to encode the variational length mel into a fix length latent representation, and then conditioned the representation to the decoder.
\item We used Gaussian-mixture-model (GMM) attention rather than location-sensitive attention.
\item We used wavenet neural vocoder~\cite{oord2016wavenet}.
\end{itemize}

For Chinese TTS, we used hundreds of speakers from internal automatic audio text processing toolkit, for English, we used libritts dataset~\cite{zen2019libritts}, and for Japanese we used JVS corpus which includes 100 Japanese speakers. As for input representations, we used phoneme with tone for Chinese, phoneme with stress for English, and phoneme with mora accent for Japanese.
In our experiment, we recorded 2 hours of Chinese voice data from an internal female speaker who speaks only Chinese for this demo.

%% file: 05avatar.tex
\begin{figure*}[ht]
    \centering
    \includegraphics[width=1.0\textwidth,keepaspectratio]
    {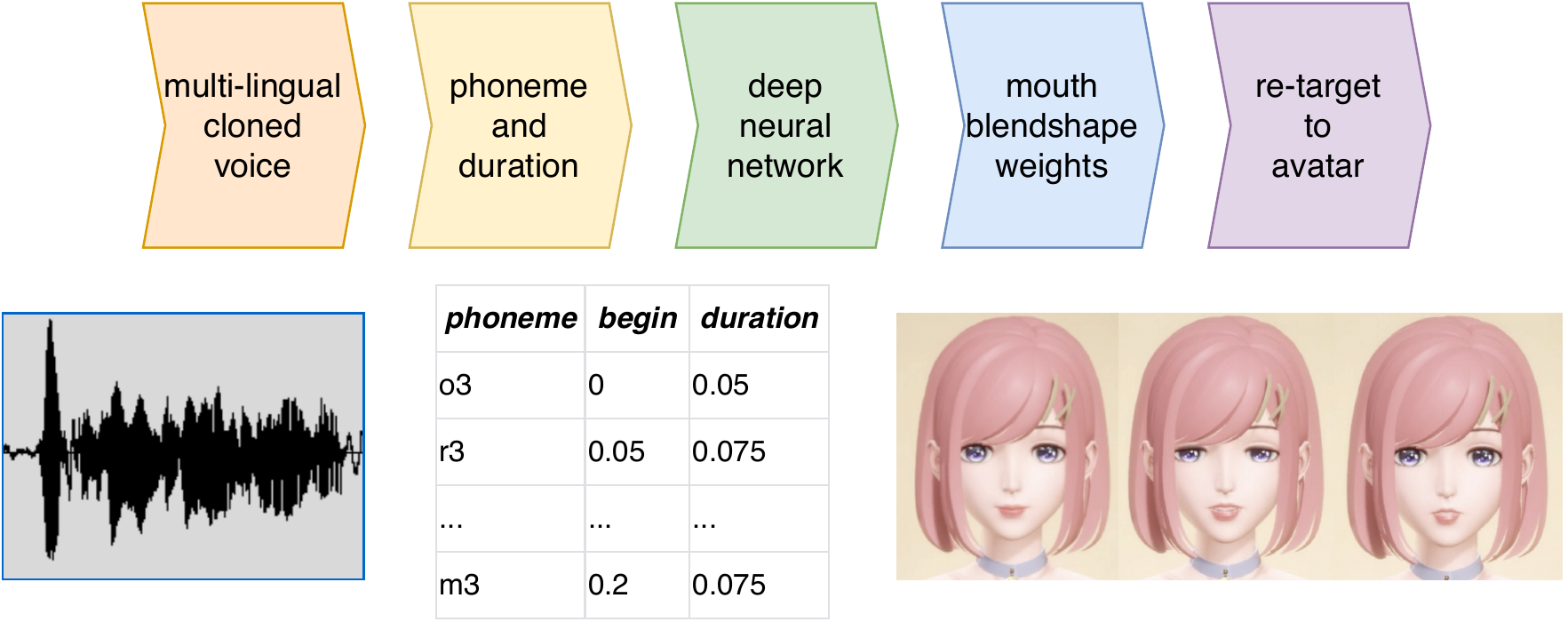}
    \caption{Avatar animation synthesis: a)~multi-lingual voices are cloned. b)~A sequence of phonemes and their duration is drawn from the voices. c)~A sequence of blendshape weights is transformed by a neural network model. d)~Lip-motion is synthesized and re-targeted synchronously to avatar animation.}
    \label{fig:acl2020/visualrending}
\end{figure*}

We believe that lifelike animated avatar will make the news broadcasting more viewer friendly.
In this section, we will describe the techniques to render the animated avatar and to synchronize the lip and facial motions. 

\subsection{Lip Syncing} 

The avatar animation module produces a set of lip motion animation parameters for each video frame, which is synced with the audio synthesized by the TTS module and used to drive the character.

Since the module should be speaker agnostic and TTS-model-independent, no audio signal is required as input. Instead, a sequence of phonemes and their duration is drawn from the TTS module and fed into the lip motion synthesis module. 
This step can be regarded as tackling a sequence to sequence learning problem. 
The generated lip motion animation parameters should  be able to be re-targeted to any avatar and easy to visualize by animators. 
To meet this requirement, the lip motion animation parameters are represented as blend weights of facial expression blendshapes. 
The blendshapes for the rendered character are designed by an animator according to the semantic of the blendshapes. In each rendered frame, the blendshapes are linear blended with the weights predicted by the module to form the final 3D mesh with correct mouth shape for rendering.

Since the module should produce high fidelity animations and run in real-time, a neural network model that has learned from real-world data is introduced to transform the phoneme and duration sequence to the sequence of blendshape weights. A sliding window neural network similar to~\citet{taylor2017deep}, which is used to capture the local phonetic context and produce smooth animations. The phoneme and duration sequence is converted to fixed length sequence of phoneme frame according to the desired video frame rate before being further converted to one-hot encoding sequence which is taken as input to the neural network in a sliding widow the length of which  is 11. Three are 32 mouth related blendshape weights  predicted for each frame in a sliding window with length of 5. Following~\citet{taylor2017deep}, the final blendshape weights for each frame is generated by blending every predictions in the overlapping sliding windows using the frame-wise mean.

The model we used is a fully connected feed forward neural network with three hidden layers and 2048 units per hidden layer. The hyperbolic tangent function is used as activation function. Batch normalization is used after each hidden layer~\cite{ioffe2015batch}. Dropout with probability of 0.5 is placed between output layer and last hidden layer to prevent over-fitting~\cite{wager2013dropout}. The network is trained with standard mini-batch stochastic gradient descent with mini-batch size of  128  and learning rate of 1e-3 for 8000 steps.

The training data is build from  3 hours of video and audio of a female speaker. Different from  ~\citet{taylor2017deep}, instead of using AAM to parameterize the face, the faces in the video frames are parameterized by fitting a blinear 3D face morphable model inspired by ~\citet{cao2013facewarehouse} built from our private 3D capture data. The poses of the 3D faces, the identity parameters and the weights of the individual-specific blendshapes of each frame and each view angle are joint solved with a cost function built from reconstruction error of the facial landmarks. The identity parameters are shared within all frames and the weights of the blendshapes are shared through view angles which have the same timestamp. The phoneme-duration sequence and the blendshape weights sequence are used to train the sliding window neural network.

\subsection{Character Rendering} 

Unity, the real time 3D rendering engine is used to render the avatar for Xiaomingbot.

For eye rendering, we used Normal Mapping to simulate the the iris, and  Parallax Mapping to simulate the effect of refraction. As for the highlights of the eyes, we used the GGX term in PBR for  approximation.
In terms of hair rendering, we used the kajiya-kay shading model to simulate the double highlights of the hair~\cite{kajiya1989rendering}, and solved the problem of translucency  using a mesh-level triangle sorting algorithm.
For skin rendering, we used the Separable Subsurface Scattering algorithm to approximate the translucency of the skin~\cite{jimenez2015separable}. 
For simple clothing materials, we used the PBR algorithm directly. For fabric and silk, we used Disney's anisotropic BRDF~\cite{burley2012physically}.

Since physical-based cloth simulation algorithm is more expensive for mobile, we used the Spring-Mass System(SMS) for cloth simulation. The specific method is to generate a skeletal system and use SMS to drive the movement of bones~\cite{liu2013fast}.
However, the above approach may cause the clothing to  overlap the body. To address this problem, we deployed some new virtual bone points to the skeletal system, and reduced the overlay using the CCD IK method~\cite{wang1991combined}, which displayed great performance in most cases.

%% file: 06conclusion.tex
In this paper, we present \method, a multilingual and multi-modal system for news reporting.
The entire process of Xiaomingbot's news reporting can be condensed as follows. First, it learns how to write news articles based on a text generation model, and summarize the news through an extraction based method. Next, its system translates the summarization into multiple languages. Finally, the system produces the video of an animated avatar reading the news with synthesized voice. Owing to the voice cloning model that can learn from a few Chinese audio samples, Xiaomingbot can maintain consistency in intonation and voice projection across different languages. So far, Xiaomingbot has been deployed online and is serving users

The system is but a first attempt to build a fully functional robot reporter capable of writing, speaking, and expressing with motion. Xiaomingbot is not yet perfect, and has limitations and room for improvement. One such important direction for future improvement is the expansion of areas that it can work in, which can be achieved through a promising approach of adopting model based technologies together with rule/template based ones. Another direction for improvement is to further enhance the ability to interact with users via a conversation interface.

%% file: paper.bbl
\begin{thebibliography}{23}
\expandafter\ifx\csname natexlab\endcsname\relax\def\natexlab#1{#1}\fi

\bibitem[{Burley and Studios(2012)}]{burley2012physically}
Brent Burley and Walt Disney~Animation Studios. 2012.
\newblock Physically-based shading at disney.
\newblock In \emph{ACM SIGGRAPH}, volume 2012, pages 1--7.

\bibitem[{Cao et~al.(2013)Cao, Weng, Zhou, Tong, and
  Zhou}]{cao2013facewarehouse}
Chen Cao, Yanlin Weng, Shun Zhou, Yiying Tong, and Kun Zhou. 2013.
\newblock Facewarehouse: A 3d facial expression database for visual computing.
\newblock \emph{IEEE Transactions on Visualization and Computer Graphics},
  20(3):413--425.

\bibitem[{Hua et~al.(2017)Hua, Wan, and Li}]{hua2017overview}
Lifeng Hua, Xiaojun Wan, and Lei Li. 2017.
\newblock \href {https://doi.org/10.1007/978-3-319-73618-1\_84} {Overview of
  the {NLPCC} 2017 shared task: Single document summarization}.
\newblock In \emph{Natural Language Processing and Chinese Computing - 6th
  {CCF} International Conference, {NLPCC} 2017, Dalian, China, November 8-12,
  2017, Proceedings}, volume 10619 of \emph{Lecture Notes in Computer Science},
  pages 942--947. Springer.

\bibitem[{Ioffe and Szegedy(2015)}]{ioffe2015batch}
Sergey Ioffe and Christian Szegedy. 2015.
\newblock \href {http://proceedings.mlr.press/v37/ioffe15.html} {Batch
  normalization: Accelerating deep network training by reducing internal
  covariate shift}.
\newblock In \emph{Proceedings of the 32nd International Conference on Machine
  Learning, {ICML} 2015, Lille, France, 6-11 July 2015}, pages 448--456.

\bibitem[{Jimenez et~al.(2015)Jimenez, Zsolnai, Jarabo, Freude, Auzinger, Wu,
  der Pahlen, Wimmer, and Gutierrez}]{jimenez2015separable}
Jorge Jimenez, K\'{a}roly Zsolnai, Adrian Jarabo, Christian Freude, Thomas
  Auzinger, Xian-Chun Wu, Javier der Pahlen, Michael Wimmer, and Diego
  Gutierrez. 2015.
\newblock \href {https://doi.org/10.1111/cgf.12529} {Separable subsurface
  scattering}.
\newblock \emph{Comput. Graph. Forum}, 34(6):188–197.

\bibitem[{Kajiya and Kay(1989)}]{kajiya1989rendering}
James~T Kajiya and Timothy~L Kay. 1989.
\newblock Rendering fur with three dimensional textures.
\newblock \emph{ACM Siggraph Computer Graphics}, 23(3):271--280.

\bibitem[{Li and Wan(2018)}]{li2018overview}
Lei Li and Xiaojun Wan. 2018.
\newblock \href {https://doi.org/10.1007/978-3-319-99501-4\_44} {Overview of
  the {NLPCC} 2018 shared task: Single document summarization}.
\newblock In \emph{Natural Language Processing and Chinese Computing - 7th
  {CCF} International Conference, {NLPCC} 2018, Hohhot, China, August 26-30,
  2018, Proceedings, Part {II}}, volume 11109 of \emph{Lecture Notes in
  Computer Science}, pages 457--463. Springer.

\bibitem[{Liu et~al.(2013)Liu, Bargteil, O'Brien, and Kavan}]{liu2013fast}
Tiantian Liu, Adam~W Bargteil, James~F O'Brien, and Ladislav Kavan. 2013.
\newblock Fast simulation of mass-spring systems.
\newblock \emph{ACM Transactions on Graphics (TOG)}, 32(6):1--7.

\bibitem[{Miao et~al.(2019)Miao, Zhou, Mou, Yan, and Li}]{miao2019cgmh}
Ning Miao, Hao Zhou, Lili Mou, Rui Yan, and Lei Li. 2019.
\newblock \href {http://arxiv.org/abs/1811.10996} {{CGMH}: Constrained sentence
  generation by metropolis-hastings sampling}.
\newblock In \emph{the 33rd {AAAI} Conference on Artificial Intelligence
  (AAAI)}.

\bibitem[{Oord et~al.(2016)Oord, Dieleman, Zen, Simonyan, Vinyals, Graves,
  Kalchbrenner, Senior, and Kavukcuoglu}]{oord2016wavenet}
Aaron van~den Oord, Sander Dieleman, Heiga Zen, Karen Simonyan, Oriol Vinyals,
  Alex Graves, Nal Kalchbrenner, Andrew Senior, and Koray Kavukcuoglu. 2016.
\newblock Wavenet: A generative model for raw audio.
\newblock \emph{arXiv preprint arXiv:1609.03499}.

\bibitem[{Radford et~al.(2019)Radford, Wu, Child, Luan, Amodei, and
  Sutskever}]{radford2019language}
Alec Radford, Jeffrey Wu, Rewon Child, David Luan, Dario Amodei, and Ilya
  Sutskever. 2019.
\newblock Language models are unsupervised multitask learners.
\newblock \emph{OpenAI Blog}, 1(8):9.

\bibitem[{Shen et~al.(2018)Shen, Pang, Weiss, Schuster, Jaitly, Yang, Chen,
  Zhang, Wang, Skerry-Ryan, Saurous, Agiomyrgiannakis, and
  Wu}]{J.Shen2018:icassp}
Jonathan Shen, Ruoming Pang, Ron~J. Weiss, Michael Schuster, Navdeep Jaitly,
  Zongheng Yang, Zhifeng Chen, Yu~Zhang, Yuxuan Wang, R.~J. Skerry-Ryan, Rif~A.
  Saurous, Yannis Agiomyrgiannakis, and Yonghui Wu. 2018.
\newblock Natural {TTS} synthesis by conditioning wavenet on {MEL} spectrogram
  predictions.
\newblock In \emph{2018 IEEE International Conference on Acoustics, Speech and
  Signal Processing (ICASSP)}, pages 4779--4783.

\bibitem[{Sun et~al.(2019)Sun, Chen, Zhou, Zhou, Li, and
  Jiang}]{sun2019graspsnooker}
Zhaoyue Sun, Jiaze Chen, Hao Zhou, Deyu Zhou, Lei Li, and Mingmin Jiang. 2019.
\newblock \href {https://doi.org/10.24963/ijcai.2019/959} {{GraspSnooker}:
  Automatic {Chinese} commentary generation for snooker videos}.
\newblock In \emph{the 28th International Joint Conference on Artificial
  Intelligence ({IJCAI})}, pages 6569--6571.
\newblock Demos.

\bibitem[{Taylor et~al.(2017)Taylor, Kim, Yue, Mahler, Krahe, Rodriguez,
  Hodgins, and Matthews}]{taylor2017deep}
Sarah Taylor, Taehwan Kim, Yisong Yue, Moshe Mahler, James Krahe,
  Anastasio~Garcia Rodriguez, Jessica Hodgins, and Iain Matthews. 2017.
\newblock A deep learning approach for generalized speech animation.
\newblock \emph{ACM Transactions on Graphics (TOG)}, 36(4):1--11.

\bibitem[{Vaswani et~al.(2017)Vaswani, Shazeer, Parmar, Uszkoreit, Jones,
  Gomez, Kaiser, and Polosukhin}]{vaswani2017attention}
Ashish Vaswani, Noam Shazeer, Niki Parmar, Jakob Uszkoreit, Llion Jones,
  Aidan~N Gomez, {\L}ukasz Kaiser, and Illia Polosukhin. 2017.
\newblock Attention is all you need.
\newblock In \emph{Advances in neural information processing systems}, pages
  5998--6008.

\bibitem[{Wager et~al.(2013)Wager, Wang, and Liang}]{wager2013dropout}
Stefan Wager, Sida Wang, and Percy~S Liang. 2013.
\newblock Dropout training as adaptive regularization.
\newblock In \emph{Advances in neural information processing systems}, pages
  351--359.

\bibitem[{Wang and Chen(1991)}]{wang1991combined}
L-CT Wang and Chih-Cheng Chen. 1991.
\newblock A combined optimization method for solving the inverse kinematics
  problems of mechanical manipulators.
\newblock \emph{IEEE Transactions on Robotics and Automation}, 7(4):489--499.

\bibitem[{Wang et~al.(2017)Wang, Skerry{-}Ryan, Stanton, Wu, Weiss, Jaitly,
  Yang, Xiao, Chen, Bengio, Le, Agiomyrgiannakis, Clark, and
  Saurous}]{wang2017tacotron}
Yuxuan Wang, R.~J. Skerry{-}Ryan, Daisy Stanton, Yonghui Wu, Ron~J. Weiss,
  Navdeep Jaitly, Zongheng Yang, Ying Xiao, Zhifeng Chen, Samy Bengio, Quoc~V.
  Le, Yannis Agiomyrgiannakis, Rob Clark, and Rif~A. Saurous. 2017.
\newblock \href
  {http://www.isca-speech.org/archive/Interspeech\_2017/abstracts/1452.html}
  {Tacotron: Towards end-to-end speech synthesis}.
\newblock In \emph{Interspeech 2017, 18th Annual Conference of the
  International Speech Communication Association, Stockholm, Sweden, August
  20-24, 2017}, pages 4006--4010.

\bibitem[{Yang et~al.(2020)Yang, Wang, Zhou, Zhao, Zhang, Yu, and
  Li}]{yang2020towards}
Jiacheng Yang, Mingxuan Wang, Hao Zhou, Chengqi Zhao, Weinan Zhang, Yong Yu,
  and Lei Li. 2020.
\newblock Towards making the most of {BERT} in neural machine translation.
\newblock In \emph{the 34th {AAAI} Conference on Artificial Intelligence
  (AAAI)}.

\bibitem[{Ye et~al.(2020)Ye, Shi, Zhou, Wei, and Li}]{ye2020variational}
Rong Ye, Wenxian Shi, Hao Zhou, Zhongyu Wei, and Lei Li. 2020.
\newblock \href {https://openreview.net/forum?id=HkejNgBtPB} {Variational
  template machine for data-to-text generation}.
\newblock In \emph{International Conference on Learning Representations
  (ICLR)}.

\bibitem[{Zen et~al.(2019)Zen, Dang, Clark, Zhang, Weiss, Jia, Chen, and
  Wu}]{zen2019libritts}
Heiga Zen, Viet Dang, Rob Clark, Yu~Zhang, Ron~J. Weiss, Ye~Jia, Zhifeng Chen,
  and Yonghui Wu. 2019.
\newblock \href {https://doi.org/10.21437/Interspeech.2019-2441} {Libritts: {A}
  corpus derived from librispeech for text-to-speech}.
\newblock In \emph{Interspeech 2019, 20th Annual Conference of the
  International Speech Communication Association, Graz, Austria, 15-19
  September 2019}, pages 1526--1530.

\bibitem[{Zhang et~al.(2016)Zhang, Yao, and Wan}]{zhang-etal-2016-towards}
Jianmin Zhang, Jin-ge Yao, and Xiaojun Wan. 2016.
\newblock \href {https://doi.org/10.18653/v1/P16-1129} {Towards constructing
  sports news from live text commentary}.
\newblock In \emph{Proceedings of the 54th Annual Meeting of the Association
  for Computational Linguistics (Volume 1: Long Papers)}, pages 1361--1371,
  Berlin, Germany. Association for Computational Linguistics.

\bibitem[{Zheng et~al.(2020)Zheng, Zhou, Huang, Li, Dai, and
  Chen}]{Zheng2020Mirror-Generative}
Zaixiang Zheng, Hao Zhou, Shujian Huang, Lei Li, Xin-Yu Dai, and Jiajun Chen.
  2020.
\newblock \href {https://openreview.net/forum?id=HkxQRTNYPH} {Mirror-generative
  neural machine translation}.
\newblock In \emph{International Conference on Learning Representations}.

\end{thebibliography}
